\documentstyle[12pt]{article}
\newcommand{\beq}{\begin{equation}}
\newcommand{\eeq}{\end{equation}}
\newcommand{\bea}{\begin{eqnarray}}
\newcommand{\eea}{\end{eqnarray}}
\newcommand{\beas}{\begin{eqnarray*}}
\newcommand{\eeas}{\end{eqnarray*}}
\newcommand{\nn}{\nonumber}

\newcommand{\limit}{\rightarrow}
\newcommand{\cN}{{\cal N}}
\newcommand{\tr}{{\rm tr}}
\newcommand{\Tr}{{\rm Tr}}
\newcommand{\bra}{\langle}
\newcommand{\ket}{\rangle}
\begin{document}
\topmargin 0pt
\oddsidemargin 5mm
\headheight 0pt
\topskip 0mm
\begin{titlepage}
\begin{flushright}
April 1999 \\
KEK-TH-622 \\
hep-th/9904122
\end{flushright}
\vspace{1cm}
\begin{Large}
\begin{center}
{\sc Cohomological Field Theory Approach} \\
{\sc to Matrix Strings} \\
\end{center}
\end{Large}
\vspace{5mm}
\begin{center}
\begin{large}
{\sc Fumihiko Sugino}\footnote{E-mail address: {\tt sugino@post.kek.jp}}\\
\end{large}
\vspace{4mm}
{\it Institute of Particle and Nuclear Studies,}\\
{\it High Energy Accelerator Research Organization (KEK),}\\
{\it Tsukuba, Ibaraki 305-0801, Japan}\\
\vspace{3cm}
\begin{large} 
Abstract 
\end{large}
\end{center}

   In this paper we consider 
IIA and IIB matrix string theories which are defined by 
two-dimensional and three-dimensional super Yang-Mills theory with the 
maximal supersymmetry, respectively. 
We exactly compute the partition function of both of the theories by 
mapping to a cohomological field theory. 
Our result for the IIA matrix string theory coincides with the result 
obtained in the infra-red limit by Kostov and Vanhove, and thus gives 
a proof of the exact quasi classics conjectured by them. 
Further, our result for the IIB matrix string theory coincides with 
the exact result of IKKT model by Moore, Nekrasov and Shatashvili. 
It may be an evidence of the equivalence between the two distinct 
IIB matrix models arising from different roots.

\end{titlepage}


\section{Introduction}

  A recent great development of nonperturbative aspects of 
superstring theory has begun with following two events. 
One is a proposal of M-theory \cite{Witten1}, which is a hypothetical 
theory describing strongly coupled region of type IIA superstring 
theory and reduces to eleven-dimensional supergravity in the low 
energy limit. 
The relation among parameters of string theory (the string length 
$l_s$ and the string coupling $g_s$) and of eleven-dimensional 
supergravity (the Planck length $l_{11}$ and 
the radius of compactified the 11-th direction $R_{11}$) is given by 
\beq
R_{11}=g_sl_s=g_s^{2/3}l_{11}.
\eeq
M-theory gives a systematic view point 
for string theory --- in considering the M-theory, 
all known perturbative 
string theories are unified via various connections from duality. 
The other is a discovery of D-branes \cite{Polchinski}, which are 
solitonic solutions of string theory and can be constructed by using 
conformal field theory in a well-defined manner. Thus they are 
powerful tools for investigating nonperturbative phenomena in 
string theory. 
Further, it was followed by a new interpretation of 
dimensional reductions of ten-dimensional supersymmetric 
Yang-Mills (SYM) theory as low energy effective theories for 
dynamics of the D-branes \cite{Witten2}. 

  Those movements culminated in the conjecture 
by Banks, Fischler, Shenker and Susskind (BFSS) \cite{BFSS}, 
that ten-dimensional SYM theory dimensionally reduced to one dimension 
which describes the low energy dynamics of D0-branes so far, in the 
infinite momentum frame, 
gives a constructive definition of the 
M-theory. This one-dimensional SYM quantum mechanical model is 
called M(atrix) theory. 
Moreover, by considering toroidal compactification on a circle and 
on a two-torus in the manner of Taylor \cite{Taylor}, 
it leads to a proposal for 
a nonperturbative definition of 
type IIA superstring theory \cite{DVV,Banks-Seiberg} and  
that of type IIB superstring theory 
\cite{Banks-Seiberg,Sethi-Susskind}, respectively.
In this paper, we call them IIA and IIB matrix string theories. 
The IIA matrix string theory is given by 
ten-dimensional SYM theory dimensionally 
reduced to two-dimensions, and the string coupling constant corresponds 
to the inverse of the SYM coupling. Also, the IIB matrix string theory 
is a dimensional reduction of the ten-dimensional SYM theory 
to three dimensions. The string coupling is given by a ratio of 
lengths of two spatial dimensions of the three-dimensional SYM theory. 
In addition, with respect to another nonperturbative definition 
of type IIB superstrings \cite{IKKT}, which is in form of 
the ten-dimensional SYM theory reduced to a point (zero dimension), 
proposed by Ishibashi, Kawai, Kitazawa and Tsuchiya, 
it is referred to IKKT model. 

  Those matrix string theories successfully reproduce some of 
known results obtained from an analysis of BPS saturated states 
consisting of fundamental strings, D-branes and their bound states 
\cite{Banks-Seiberg-Shenker}. 
This fact is considered as one of evidences 
that the matrix string theories truely are constructive 
definitions of string theories. However, with respect to 
nonperturbative dynamics of the string theories, in order to 
investigate it we cannot help treating nonperturbative dynamics of 
the SYM theory sides. 
Though understanding of this area is now 
in development \cite{Seiberg,PSS}, 
unfortunately at present it seems to be not powerful enough for 
handling the problem. 

   In this paper, we consider the partition function of the matrix 
string theories --- it is one of the most fundamental quantities 
reflecting dynamical property of vacua of the theories. 
The partition function of the IIA matrix string theory was computed 
in strongly coupled limit of the SYM theory 
by Kostov and Vanhove \cite{Kostov-Vanhove}. 
For the partition function of the IKKT model, the exact result was 
obtained by interpreting the theory 
as a cohomological field theory 
by Moore, Nekrasov and Shatashvili \cite{Moore-Nekrasov-Shatashvili}. 
Here, we exactly calculate the partition function of 
both of IIA and IIB 
matrix string theories by mapping the theories to 
cohomological field theories. 
In the IIA case our result coincides with the result obtained in the 
strongly coupled limit, which thus gives a proof of 
{\it exact quasi classics} 
discussed in \cite{Kostov-Vanhove}. 
Also in the IIB case, our result agrees with the exact calculation for 
the IKKT model, which seems to suggest the 
equivalence between those two different nonperturbative formulations 
of IIB string theory.  

   The paper is organized as follows. In the next section, we introduce 
the IIA and IIB matrix string theories via toroidal compactification   
of the M(atrix) theory. We briefly explain how string interactions 
emerge from the SYM theory 
and show how spinors with correct chirality appear. 
In section 3, as a preparation for computing the partition function of 
the matrix strings, we evaluate the partition function of 
four-dimensional $\cN=4$ $SU(N)$ SYM theory defined on a four-torus 
by mapping the theory to a cohomological field theory via 
a twisting procedure. 
Then in section 4, we exactly compute the partition function of the IIA 
matrix string theory, and see that our result coincides with the result 
in the infra-red limit by Kostov and Vanhove. 
In section 5, for the partition function of the IIB matrix string theory 
we perform the calculation in the ten-dimensional IIB limit, and find 
the identical result with the partition function of the IKKT model 
by Moore, Nekrasov and Shatashvili. 
Finally, section 6 is devoted to conclusions. 
In Appendix A, we clarify the connection between two-dimensional 
$\cN=2$ SYM theory and a cohomological field theory used in section 4.


\section{Matrix Strings}

  In this section we review some basic properties of the IIA and IIB 
matrix string theories, which are derived from the M(atrix) theory by 
considering toroidal compactification on $S^1$ and on $T^2$ 
respectively. In addition, we give an argument 
that two spinors of the same space-time chirality emerge 
in the IIB matrix string theory, which has not been found 
in the literatures. 

   We start with the M(atrix) theory \cite{BFSS}, whose action 
has the same form as low energy effective action of $N$ D0-branes, 
that is one-dimensional $U(N)$ SYM theory with 16 supercharges. 
The BFSS conjecture is that this action exactly describes M-theory in 
the decompactified limit $R_{11} \limit \infty$ by going to 
the infinite momentum frame. The infinite momentum frame 
means infinite amount of boosting along the 11th direction, 
i.e. momentum of the 11th direction becomes 
\beq
p_{11}=\frac{N}{R_{11}}\limit \infty.
\eeq 
At the same time we must take the limit 
$N,\; R_{11} \limit \infty,$ in order that the resulting theory 
represents the strongly 
coupled limit of type IIA superstring theory.

\subsection{IIA Matrix String Theory}

  Let us consider the compactification of one of the transverse 
directions, say the 9th, to a circle of the circumference $L,$ 
denoted by $S^1(L).$ 
Since in the M(atrix) theory we consider the limit that 
the 11th direction is decompactified, 
we have ten-dimensional 
type IIA string theory with the string length 
\beq
l_s^2=\frac{2\pi l_{11}^3}{L}. 
\eeq
According to the prescription of Taylor \cite{Taylor}, 
this IIA string theory is described by two-dimensional $\cN =8 \; 
U(N)$ SYM theory, which is a dimensional reduction of 
ten-dimensional $\cN=1$ SYM theory \cite{DVV}, 
\begin{eqnarray}
S_{{\rm IIA-MS}} & = & \frac{1}{g^2}\int dt \int_0^Rd\sigma\;\tr
\left[-\frac 14F_{\mu\nu}F^{\mu\nu}-\frac 12(D_{\mu}X^I)^2
+i\theta^T(D_t+\Gamma^9D_{\sigma})\theta \right.\nn \\
  &  & \left. +\frac 14 [X^I, X^J]^2
              +\theta^T\Gamma^I[X_I,\theta]\right]. 
\label{actionIIA-MS}
\end{eqnarray}
Here, $F_{\mu\nu}=\partial_{\mu}A_{\nu}-\partial_{\nu}A_{\mu}
-i[A_{\mu},A_{\nu}]$ is a field strength made from two-dimensional 
gauge field $A_{\mu},$  
$X^I$ ($I=1,\cdots,8$) are Higgs fields and $\theta^{\alpha}$ 
($\alpha=1,\cdots,16$) are fermions. 
We use the convention that 
the $\Gamma$-matrices are $16\times 16$ real symmetric matrices 
satisfying 
\beq
\{\Gamma^I,\Gamma^J\}=2\delta^{IJ}, \hspace{1cm} 
\Gamma^9=\Gamma^1\cdots\Gamma^8.
\eeq
$\theta^{\alpha}$'s are decomposed into the spinor and conjugate 
spinor representations (${\bf 8_s}\oplus {\bf 8_c}$) 
of the rotational group in the transverse directions $SO(8),$ 
which are characterized by the eigenvalues of $\Gamma^9,$ 
so called chirality: 
\beq
\theta^{\alpha}=\theta^{\alpha}_++\theta^{\alpha}_-, \hspace{1cm}
\Gamma^9\theta^{\alpha}_{\pm}=\pm\theta^{\alpha}_{\pm}, 
\label{chiraldecomposition}
\eeq
and $X^I$'s transform as the vector ${\bf 8_v}.$ 
The spatial coordinate $\sigma,$ which arises as performing the 
compactification, takes a value on the circle dual to 
$S^1(L)$: $0\leq\sigma\leq R,$ where\footnote{
The relation between parameters in the matrix string theory and those 
in SYM theory, eqs. (\ref{RLrelation}) and (\ref{IIAggs}) in the IIA case 
as well as eqs. (\ref{*}), (\ref{**}) and (\ref{IIBcoupling}) 
in the IIB case, can be derived either by tracing the procedure by 
Taylor \cite{Taylor} or 
by employing another argument in ref. \cite{FHRS}.} 
\beq
R=(2\pi)^2\frac{l_s^2}{L}.
\label{RLrelation}
\eeq
The coupling constant of the SYM theory $g$ is related to the string 
coupling $g_s$ as 
\beq
g^2=\frac{2\pi}{(Rg_s)^2}.
\label{IIAggs}
\eeq

  Weakly coupled strings are recovered by considering the limit 
$g_s\limit 0$ or equivalently the infra-red limit of the 
SYM theory \cite{DVV}. In this situation, the theory is 
described by the eigenvalues of the simultaneously 
diagonalizable configurations of $X^I$ and $\theta^{\alpha}$: 
\beq
\tilde{X^I} = {\rm diag}(x^I_1,\cdots,x^I_N), \hspace{1cm}
\tilde{\theta^{\alpha}} = 
  {\rm diag}(\theta^{\alpha}_1,\cdots, \theta^{\alpha}_N)
\eeq
where 
\beq
X^I=V\tilde{X^I}V^{\dagger}, \hspace{1cm}
\theta^{\alpha}=V\tilde{\theta^{\alpha}}V^{\dagger}. 
\label{diagonalpart}
\eeq
The angular variables $V\in U(N)$ and nontrivial 
configurations of the gauge field yielding 
non-zero curvature $F_{\mu\nu}$ are 
energetically decoupled from the theory in the infra-red limit, 
and the action (\ref{actionIIA-MS}) reduces to the action of 
multiple Green-Schwarz superstrings in the light-cone gauge. 
Then the gauge field can take the pure gauge configuration 
\beq
A_{\mu}=iV\partial_{\mu}V^{\dagger}.
\eeq
Also, the $x^I_i$ and $\theta^{\alpha}_i$ can represent 
strings of various lengths by considering the multi-valued 
configuration: 
\begin{eqnarray}
 & & \tilde{X^I}(t,\sigma+R) =
        g^{\dagger}\tilde{X^I}(t,\sigma)g, \hspace{1cm}
\tilde{\theta^{\alpha}}(t,\sigma+R) =
         g^{\dagger}\tilde{\theta^{\alpha}}(t,\sigma)g, \nn \\
 & & V(t,\sigma+R) = V(t,\sigma)g, 
\label{BCofdiagonalpart}
\end{eqnarray}
where $g$ is an element of the Weyl group of $U(N),$ i.e. the 
permutation group $S_N,$ which permutes the $N$ eigenvalues. 
In going around the $\sigma$-direction, the eigenvalues are 
interchanged by the action of $g$ in eqs. (\ref{BCofdiagonalpart}), 
and as a result they form cycles of various 
lengths corresponding to permutation cycles in $g.$ 
Each cycle is interpreted as a single closed string, and thus 
for $g\in S_N$ 
consisting of $N_n$ $n$-cycles 
(satisfying $N=\sum_nnN_n$) one has $N_n$ strings with length 
$n.$ 

  Note that the total matrices  
$X^I,$ $\theta^{\alpha}$ 
in eqs. (\ref{diagonalpart}) and the gauge field $A_{\mu},$ 
which appear in the SYM theory, 
remain single-valued although we consider 
the case that the variables of string coordinates 
$\tilde{X^I}$ and $\tilde{\theta^{\alpha}}$ are multi-valued 
as in eqs. (\ref{BCofdiagonalpart}), 
which is argued in ref. \cite{Wynter}. 
For the simplest 
$N=2$ case, the multi-valued configuration is represented by 
\beq
\tilde{X^I}=\left(\begin{array}{cc} x^I_1 & 0 \\
                                    0 & x^I_2
                  \end{array}\right),  \hspace{1cm}
\tilde{\theta^{\alpha}}=\left(\begin{array}{cc} 
                                \theta^{\alpha}_1 & 0 \\
                                0 & \theta^{\alpha}_2
                  \end{array}\right),  \hspace{1cm}
V=\frac{1}{\sqrt{2}}\left(\begin{array}{cc}
                   e^{i\pi\sigma/R} & -e^{i\pi\sigma/R} \\
                            1 & 1 
                     \end{array}\right),
\eeq
where 
$$
x^I_1(t,\sigma+R)=x^I_2(t,\sigma), \hspace{1cm}
x^I_2(t,\sigma+R)=x^I_1(t,\sigma),
$$
and $\theta^{\alpha}_i$ satisfies the same boundary condition. 
The matrices $X^I,$ $\theta^{\alpha}$ and $A_{\mu}$ are 
single-valued. In particular, the gauge field is  
\beq
A_t=0, \hspace{1cm} 
A_{\sigma}=\frac 12 \left(\begin{array}{cc} 1 & 0 \\
                                            0 & 0 
                          \end{array}\right),
\eeq
which cannot be eliminated by any single-valued gauge 
transformation. 

 As discussed in \cite {DVV}, 
relaxing the strict limit $g_s=0,$ the number of the 
strings are no longer conserved. We can interpret 
this as a result of string interactions 
(splitting or joining of one or two strings) which occur 
when the two sets of eigenvalues $\{x^I_i\}_{I=1,\cdots,8}$ and 
$\{x^I_j\}_{I=1,\cdots,8}$ coincide. 
Whenever the interaction occurs, the boundary condition of 
$\tilde{X^I},$ $\tilde{\theta^{\alpha}}$ and $V$ is changed, 
i.e. a branch point appears on a cylinder coordinated by 
$t$ and $\sigma.$ 
We should remark that as discussed in \cite{GHV}, 
in spite of this singular configuration 
of the diagonal variables, for a neighborhood of the
branch point the corresponding total matrices 
exist as a smooth and single-valued 
configuration of the SYM theory. 
In the section 4 
of ref. \cite{GHV} such a configuration is constructed 
in the $N=2$ case, which is 
given by a solution of the dimensionally reduced version 
of the self-dual equation of four-dimensional Yang-Mills 
theory to two dimensions: 
\begin{eqnarray}
 & & [X_1,X_2]=\frac{i}{g^2}F_{t\sigma}, \label{selfdualeq} \\
 & & D_tX_1=-D_{\sigma}X_2, \nn \\
 & & D_tX_2=D_{\sigma}X_1,
\end{eqnarray}
and the all other fields are set to zero.  
Compared to the configuration in the strict limit ($g^2=\infty$) 
that is $[X_1,X_2]=0,$ the $O(1/g^2)$-correction in 
eq. (\ref{selfdualeq}) works well to make the configuration 
in the SYM theory smooth\footnote{An extension of this argument 
in the case of general $N$ is discussed in refs. \cite{BBN}.
I thank L. Bonora for informing me of those literatures.}.

\subsection{IIB Matrix String Theory} 

   When compactifying IIA string theory to a circle and taking 
T-dual to the circle, we have IIB string theory on $S^1.$ 
In the limit of shrinking the circle of the IIA theory, the $S^1$ of 
the IIB theory is decompactified, thus we have IIB string theory in 
ten dimensions. From the view point of M-theory, IIA theory 
on $S^1$ means M-theory on $T^2.$ We denote the size of the 
two-torus by $L_1$ and $L_2,$ i.e. $T^2=S^1(L_1)\times S^1(L_2).$ 
Since there is no distinctive meaning between the two $S^1$'s, 
now we have 
two ways to obtain IIB theory 
corresponding to shrinking either $L_1$ or 
$L_2.$ From the analysis based on low energy effective 
theories, two IIB theories we obtain as the result are believed to be 
equivalent and connected by S-duality \cite{Aspinwall,Schwarz}. 
In this way, the M-theory perspective yields a geometrical 
interpretation to the S-duality in type IIB string theory.

  Let us consider this operation in the M(atrix) theory. 
Then we obtain three-dimensional $\cN=8$ $U(N)$ SYM theory, 
which is dimensionally reduced from the ten-dimensional theory, as 
a matrix model corresponding to IIB theory on $S^1$ 
\cite{Banks-Seiberg,Sethi-Susskind}: 
\begin{eqnarray}
S_{{\rm IIB-MS}} & = & \frac{1}{g^2}\int dt \int_0^{R_1}d\sigma_1
\int_0^{R_2}d\sigma_2\;\tr
\left[-\frac 14F_{\mu\nu}F^{\mu\nu}-\frac 12(D_{\mu}X^I)^2 
\right.\nn \\
 & & \left.+i
      \theta^T(D_t+\Gamma^8D_{\sigma_1}+\Gamma^9D_{\sigma_2})
       \theta  
       +\frac 14 [X^I, X^J]^2
              +\theta^T\Gamma^I[X_I,\theta]\right]. 
\label{actionIIB-MS}
\end{eqnarray}
Here we compactified the 8th and 9th transverse directions to 
the above mentioned (rectangular) two-torus. The size of the 
spatial directions ($\sigma_1$ and $\sigma_2$)  
which the SYM theory is defined on is related 
by T-duality to the two-torus, 
\beq
R_i=(2\pi)^2\frac{l_s^2}{L_i} \hspace{1cm} (i=1,2), 
\label{*}
\eeq
and the SYM coupling are given by 
\beq
g^2=\frac{R_1R_2R_{11}}{(2\pi)^2l_s^4}.
\label{**}
\eeq
It is remarkable that the string coupling is given by 
a ratio of the two 
lengths of the torus:
\beq
g_s=\frac{R_1}{R_2}=\frac{L_2}{L_1}, 
\label{IIBcoupling}
\eeq
which gives the geometrical understanding of S-duality 
also in the matrix string level. 
In the vanishing torus limit $L_1, L_2\limit 0,$ a new dimension 
opens up and becomes 
decompactified, so this limit with the ratio $L_2/L_1$ 
fixed is considered to give type IIB theory of the coupling $g_s$ 
determined by eq. (\ref{IIBcoupling}) \cite{Sethi-Susskind,FHRS}.
Note that the abelian part of 
the field strength $F_{\mu\nu}$ made from 
three-dimensional gauge field $A_{\mu}$ reduces to a single 
scalar field via duality transformation. 
Although the manifest symmetry of the action (\ref{actionIIB-MS}) is 
$SO(7),$ it is considered that 
this scalar and the seven Higgs fields $X^I$ 
($I=1,\cdots,7$) 
together belong to ${\bf 8_v}$ in $SO(8)$ 
in the ten-dimensional IIB limit 
\beq
R_1R_2\limit \infty,   \hspace{1cm}   g_s :{\rm fixed}
\label{10DIIBlimit}
\eeq 
by the argument for BPS states \cite{FHRS} and by the analysis 
of the moduli space of the three-dimensional SYM theory 
\cite{Banks-Seiberg,Seiberg}.
The fermions $\theta^{\alpha}$ ($\alpha=1,\cdots,16$) are 
to represent 
two space-time spinors of the same chirality in 
the IIB theory. In fact, after decomposing by the eigenvalues of 
$\Gamma^9$ as in eq. (\ref{chiraldecomposition}), we put 
\beq
\psi^{\alpha}_+=\Gamma^8\theta^{\alpha}_-.
\label{psi+}
\eeq
Then fermion part of the lagrangian density takes the form 
\begin{eqnarray}
 & & \tr \;[i\theta^T_+(D_t+D_{\sigma_2})\theta_+
          +i\psi^T_+(D_t-D_{\sigma_2})\psi_+ 
          +i\theta^T_+D_{\sigma_1}\psi_+
          +i\psi^T_+D_{\sigma_1}\theta_+  \nn \\
 & & \; +\theta^T_+\Gamma^I\Gamma^8[X^I,\psi_+]
        -\psi^T_+\Gamma^I\Gamma^8[X^I,\theta_+]]. 
\label{IIBfermionpart}
\end{eqnarray}
Both of $\theta_+$ and $\psi_+$ are spinors with respect to 
the manifest symmetry $SO(7)$ 
of the same eigenvalue (+1) of $\Gamma^9.$ 
In the weakly coupled limit $g_s\ll 1,$ which means $R_1\ll R_2,$ 
nonzero modes of $\partial_{\sigma_1}$ 
energetically decouple, and $\theta_+$ and $\psi_+$ represent 
the two ${\bf 8_s}$ spinors in IIB theory. 
Also, in the strongly coupled limit $R_2\ll R_1$, 
it turns out that 
$\xi_+=(\theta_++\psi_+)/\sqrt 2$ and 
$\eta_+=(\theta_+-\psi_+)/\sqrt 2$ become spinors desired in 
IIB theory\footnote{
In IIA matrix string theory, 
by the same replacement (\ref{psi+}), we seem to have 
chiral IIB theory, but it is not correct. 
The IIA matrix string theory has the manifest symmetry $SO(8).$ 
Note that $\psi_+$ is not a spinor under the $SO(8)$ due to 
the $\Gamma^8$ factor in eq. (\ref{psi+}).}. 
In this way, we can see that 
in both of the two limits related via S-duality chiral spinors are 
correctly 
reproduced in the IIB matrix string theory.


\section{Partition Function of four-dimensional $\cN=4$ $SU(N)$ 
Super Yang-Mills Theory}

  Here we calculate the partition function of 
$\cN=4$ $SU(N)$ SYM theory on four-torus $T^4$ 
by mapping the theory to a cohomological field theory. 
The dimensional reduced version of this argument 
is used in later computations 
of the partition function of IIA and IIB matrix string theories. 
  
  $\cN=4$ $SU(N)$ SYM theory on a flat four-dimensional space-time 
is given by the following lagrangian density 
defined on the $\cN=1$ 
superspace $(x,\theta,\bar{\theta})$: 
\begin{eqnarray}
{\cal L}_{\cN=4} & = & \frac{1}{16g^2}\tr 
      \left(W^{\alpha}W_{\alpha}|_{\theta\theta}+
         \bar{W}_{\dot{\alpha}}
         \bar{W}^{\dot{\alpha}}|_{\bar{\theta}\bar{\theta}}\right)
                                 \nn \\
  &  & +\frac{1}{g^2}\tr \left(\Phi_1^{\dagger}e^V\Phi_1+
                          \Phi_2^{\dagger}e^V\Phi_2+
                           \Phi_3^{\dagger}e^V\Phi_3\right)
               |_{\theta\theta\bar{\theta}\bar{\theta}} \nn \\
  &  &  +\frac{1}{\sqrt{2}g^2}\tr \left( 
       \Phi_1[\Phi_2,\Phi_3]|_{\theta\theta}+
      \Phi_1^{\dagger}[\Phi_3^{\dagger},\Phi_2^{\dagger}]
           |_{\bar{\theta}\bar{\theta}}\right). 
\label{4DN=4}
\end{eqnarray}
In this section, we use the notation in ref. \cite{Wess-Bagger}. 
The vector superfield $V$ represents a multiplet containing 
a gauge field $A_m$ and 
a complex two-component gauge fermion $\lambda,$ 
and the chiral superfield 
$\Phi_s$ ($s=1,2,3$) contains a multiplet of a complex Higgs scalar 
$B_s$ and a Higgsino $\psi_s.$ 
All the fields belong to the adjoint representation of $SU(N),$ 
``$\tr$'' denotes the trace in the fundamental representation. 
The theory has the internal symmetry group $SU(4)_I,$ 
under which the fermions ($\lambda$ and $\psi_s$'s) transform 
together as {\bf 4} and $B_s$'s as {\bf 6}. 
Also, the supercharges $Q_{\alpha}^v,$ $\bar{Q}_{v \dot{\alpha}}$ 
as {\bf 4}, ${\bf \bar{4}}$ respectively, where 
$v\;(=1,\cdots,4)$ is the $SU(4)_I$ index and 
$\alpha,$  $\dot{\alpha}$ are Lorentz indices belonging to 
$SU(2)_L,$ $SU(2)_R.$ 

   In terms of the component fields, the lagrangian takes the form 
\begin{eqnarray}
{\cal L}_{\cN=4} & = & \frac{1}{g^2}\tr \left[
-\frac 14 F^{mn}F_{mn}-i\bar{\lambda}\bar{\sigma}^mD_m\lambda
\frac 12 D^2\right] \nn \\
 & & \left.+\frac{1}{g^2}\sum_{s=1}^3\tr\right[
      F_s^{\dagger}F_s-(D^mB_s)^{\dagger}(D_mB_s)
        -i\bar{\psi}_s\bar{\sigma}^mD_m\psi_s \nn \\
 & & \hspace{15mm} 
       \left. -\frac{i}{\sqrt 2}B_s^{\dagger}[\lambda,\psi_s]
        +\frac{i}{\sqrt 2}B_s[\bar{\lambda},\bar{\psi}_s]
        +\frac 12 D[B_s,B_s^{\dagger}]\right] \nn \\
 & & +\frac{1}{\sqrt 2 g^2}\tr (F_1[B_2,B_3]+F_2[B_3,B_1]+F_3[B_1,B_2]
                                \nn \\
 & & \hspace{15mm} 
       -B_1[\psi_2,\psi_3]-B_2[\psi_3,\psi_1]-B_3[\psi_1,\psi_2]
                                \nn \\
 & & \hspace{15mm}  +{\rm h. c.}), 
\end{eqnarray}
where 
$F_s$ and $D$ are auxiliary fields appearing in $\Phi_s$ and $V.$

\subsection{Twisting of $\cN=4$ Super Yang-Mills Theory}

  The twisting procedure, which was first introduced by Witten 
\cite{Witten3}, gives a systematic tool for 
constructing a cohomological 
field theory from the original physical theory. 
Here we briefly explain a twisting procedure of $N=4$ SYM theory 
adopted by 
Vafa and Witten \cite{Vafa-Witten}. 

  First, we consider a general four-manifold on which the $\cN=4$ 
SYM theory is defined. The holonomy group of this manifold is 
$SO(4)=SU(2)_L\times SU(2)_R,$ that is a gauged symmetry with the 
spin connection being the (external) gauge field. 
Adding the global $SU(4)_I,$ we concentrate 
the symmetry group of the theory 
$$
H=SU(2)_L\times SU(2)_R\times SU(4)_I.
$$
The twisting by Vafa and Witten is performed as follows. 
Consider the subgroup of the internal symmetry group $SU(4)_I$: 
$$
SU(4)_I\supset SO(4)=SU(2)_F\times SU(2)_{F'}, 
$$
then replace the action of $SU(2)_L$ by the diagonal sum 
$SU(2)'_L=SU(2)_L\oplus SU(2)_{F'}.$ 
Thus the twisted symmetry group becomes 
$$
H'=SU(2)'_L\times SU(2)_R\times SU(2)_F.
$$
Under $H',$ the supercharges split up as 
\begin{eqnarray*}
 & & Q_{\alpha}^v=Q_{\alpha}^{ij}\limit 
      Q_{\alpha}^{i\beta}=\left\{\begin{array}{l}
            Q_{\alpha}^{i\alpha}\equiv Q^i \\
      \frac 12 \varepsilon_{\gamma (\beta}Q_{\alpha)}^{i\gamma}
                      \equiv Q_{\alpha\beta}^i
             \end{array}\right. \\
 & & \bar{Q}_{v \dot{\alpha}}=\bar{Q}_{ij\dot{\alpha}}\limit 
                     \bar{Q}_{i\beta\dot{\alpha}}, 
\end{eqnarray*}
where $i$ and $j$ are $SU(2)_F$ and $SU(2)_{F'}$ indices 
respectively, also $j$ is converted to $\beta$ by the twisting. 
Here we have the two scalar supercharges $Q^i$ ($i=1,2$), which 
are nilpotent 
\beq
\{Q^i, Q^j\}=0
\eeq
as seen from the supersymmetry algebra of the original 
supercharges  
\begin{eqnarray}
 & & \{Q_{\alpha}^v,\bar{Q}_{w \dot{\alpha}}\}=
       2\sigma^m_{\alpha\dot{\alpha}}\delta^v_w P_m, \nn \\
 & & \{Q_{\alpha}^v,Q_{\beta}^w\}=
       \{\bar{Q}_{v\dot{\alpha}},\bar{Q}_{w\dot{\beta}}\}=0. 
\label{***}
\end{eqnarray}
We can construct a cohomological field theory based on 
each $Q^i$'s. The $Q^i$'s are related to the original supercharges as 
\begin{eqnarray}
Q^1 & = & Q_1^{11}+Q_2^{12}=Q_{\alpha=1}^{v=1}+Q_{\alpha=2}^{v=2}, 
                    \nn \\
Q^2 & = & Q_1^{21}+Q_2^{22}=Q_{\alpha=1}^{v=3}+Q_{\alpha=2}^{v=4}. 
\label{Qrelation}
\end{eqnarray}

  Next, in the case of four-dimensional K\"{a}hler manifolds, 
the number of the nilpotent scalar supercharges become doubled 
as we will see\footnote{This property was explored first by Witten 
\cite{Witten4} in the case of $\cN=2$ SYM theory.}. 
A four-dimensional K\"{a}hler manifold has the reduced 
holonomy group $U(1)_L\times SU(2)_R,$ where $U(1)_L$ is a certain 
subgroup of $SU(2)_L.$ In this case, twisting is done by replacing 
the $U(1)_L$ by the diagonal sum 
$U(1)'_L=U(1)_L\oplus U(1)_{F'}.$ 
Here, $U(1)_{F'}$ is a subgroup of $SU(2)_{F'}.$ 
We assign the $U(1)_L$ charge $+1$ to $Q_{\alpha=1}^v$ and $-1$ 
to $Q_{\alpha=2}^v,$ and give a similar assignment of the 
$U(1)_{F'}$ charge ($+1$ to the $j=1$ index and $-1$ to the $j=2$). 
Then by twisting, the four $Q_{\alpha}^v$'s 
in eqs. (\ref{Qrelation}) are to have zero $U(1)'_L$ charge, that is, 
they all become scalar supercharges, since 
$Q_1^{i1}=Q_{12}^i$ and $Q_2^{i2}=-Q_{21}^i.$ 
It is clear that they are nilpotent. 
Note that one of the four charges is a (chiral) generator of 
the manifest $\cN=1$ supersymmetry in eq. (\ref{4DN=4}). 
We will use that charge, denoted by $Q,$ to construct 
a cohomological field theory in the next subsection. 

   Since the twisting changes the coupling of the fields with 
the spin connection \cite{Lab-Marino1}, 
for a general four-manifold the twisted theory 
differs from the original theory. However, on a manifold 
with the flat metric ($T^4$) or more generally 
on a hyper-K\"{a}hler manifold for which the twisting is a trivial 
operation 
owing to its holonomy group $SU(2)_R,$ the twisted theory coincides 
with the original theory. Now we will investigate the $T^4$ case. 

\subsection{Partition function of $\cN=4$ SYM on $T^4$}

   The $\cN=4$ $SU(N)$ SYM theory defined on $T^4$ 
can be seen as a cohomological field theory by (trivial) twisting, 
because of the reason mentioned above. 
In the action $S_{\cN=4}$ (\ref{4DN=4}), since $Q$ acts as 
$\frac{\partial}{\partial\theta^{\alpha}}=\int d\theta^{\alpha}$ 
(up to total derivative terms) for suitable $\alpha,$ 
the term with $|_{\theta\theta}$ are written in the $Q$-exact form 
$\{Q,\cdots\}.$ Also, the term $\tr\bar{W}_{\dot{\alpha}}
         \bar{W}^{\dot{\alpha}}|_{\bar{\theta}\bar{\theta}}$ 
is equal to $\tr W^{\alpha}W_{\alpha}|_{\theta\theta}$ modulo 
the total derivative term $\tr F\wedge F,$ 
so it is of the $Q$-exact form. 
Here we consider the periodic boundary condition 
case on $T^4,$ or equivalently the sector of zero 't Hooft 
discrete magnetic flux, those total derivative terms can be 
discarded. 
The term 
$
\tr \;\Phi_1^{\dagger}[\Phi_3^{\dagger},\Phi_2^{\dagger}]
           |_{\bar{\theta}\bar{\theta}}
$ 
can not be written as the $Q$-exact form, 
but it is $Q$-invariant because $Q$ is a 
part of the manifest $\cN=1$ supersymmetric transformation. 
Thus the action 
can be written as 
\beq
S_{\cN=4}=\{Q,\cdots\}+{\cal O}^{(0)}, 
\eeq
where 
\beq
{\cal O}^{(0)}= \int d^4x \frac{1}{\sqrt 2g^2}\tr \; 
\Phi_1^{\dagger}[\Phi_3^{\dagger},\Phi_2^{\dagger}]
           |_{\bar{\theta}\bar{\theta}} 
\eeq
is a $Q$-invariant operator and the superscript 0 stands for 
the $U(1)$-charge defined by the rotation 
\begin{eqnarray}
\Phi_s(\theta,\bar{\theta},x) & \limit & e^{\frac 23i\beta}
      \Phi_s(e^{-i\beta}\theta,e^{i\beta}\bar{\theta},x) \nn \\
W^{\alpha}(\theta,\bar{\theta},x) & \limit & 
  e^{i\beta}W^{\alpha}(e^{-i\beta}\theta,e^{i\beta}\bar{\theta},x)
\label{U(1)charge}
\end{eqnarray}
with $\beta$ being a real parameter. 
The charge of the component fields can be read off from eqs. 
(\ref{U(1)charge}) as 
$-\frac 23,\;+\frac 13,\;+\frac 43,\;0,\;-1,\;0$ for 
$B_s,\;\psi_s,\;F_s,\;A_m,\;\lambda,\;D,$ respectively. 
Note that the action has the symmetry under this rotation. 

  Now the problem can be made more tractable by considering the following 
mass perturbation 
\beq
\Delta S = \int d^4x \frac{1}{\sqrt 2 g^2} \left(-\frac m2\right) \tr \; 
[(\Phi_1^2+\Phi_2^2+\Phi_3^2)|_{\theta\theta}
 +(\Phi_1^{\dagger 2}+\Phi_2^{\dagger 2}+\Phi_3^{\dagger 2})
|_{\bar{\theta}\bar{\theta}}].
\label{masspert}
\eeq
Using the same argument as above, 
the first term is written in the form $\{Q,\cdots\},$ and the 
second term is a $Q$-invariant operator with the $U(1)$-charge 
$-\frac 23$ denoted by ${\cal O}^{(-2/3)}.$ 
In cohomological field theories, the perturbation 
by the $Q$-exact operators does not alter the theory, 
and the partition function is invariant under the perturbation of 
${\cal O}^{(-2/3)}$ if the $U(1)$-symmetry remains left even 
in the quantum level. In fact, it is so since there is no anomaly 
in $\cN=4$ theory. Thus we can obtain the answer 
by computing the partition function of 
the mass perturbed system. 

   Here, it is remarked that the 
$U(1)$-symmetry is different from the ghost number symmetry usually 
used in cohomological field theories. In fact, it is seen that 
in the cohomological field theory 
made from $\cN=4$ SYM theory the ghost number symmetry corresponds to 
the following rotation \cite{Vafa-Witten,Lab-Marino2}
\beas
\Phi_1(\theta,\bar{\theta},x) & \limit & 
  \Phi_1(e^{-i\beta}\theta,e^{i\beta},x) \\
\Phi_2(\theta,\bar{\theta},x) & \limit & 
  \Phi_2(e^{-i\beta}\theta,e^{i\beta},x) \\
\Phi_3(\theta,\bar{\theta},x) & \limit & 
  e^{2i\beta}\Phi_3(e^{-i\beta}\theta,e^{i\beta},x) \\
W^{\alpha}(\theta,\bar{\theta},x) & \limit & 
  e^{i\beta}W^{\alpha}(e^{-i\beta}\theta,e^{i\beta}\bar{\theta},x).
\eeas
Vafa and Witten \cite{Vafa-Witten} discussed the invariance of the 
partition function under 
the mass perturbation based on this ghost number symmetry. 
Our argument presents another proof of the invariance.

  We should remark that cohomological field theories have an important 
feature that the contribution of the path integration localizes 
in configurations of the classical vacua. In the mass perturbed system, 
the classical vacua are given by the solutions of these equations: 
\begin{eqnarray}
 & &    [ B_1, B_2]=m B_3, \nn \\ 
 & &    [ B_2, B_3]=m B_1, \nn \\ 
 & &    [ B_3, B_1]=m B_2     
\label{classicalvacua1}
\end{eqnarray}
with 
\beq
\sum_{s=1}^3[B_s,B_s^{\dagger}]=0. 
\label{classicalvacua2}      
\eeq
A special property of the perturbation is that three equations in 
(\ref{classicalvacua1}) have the form of $SU(2)$ algebra. 
So, the solution $\frac im B_s$ is given by the generator of 
$N$-dimensional representation 
of $SU(2).$ At the same time, this solution satisfies eq. 
(\ref{classicalvacua2}) also, because $B_s$'s are anti-hermitian. 
Since we must consider the various $N$-dimensional representations, 
both of reducible and irreducible, in general the solution takes the 
form 
\beq
\frac im B_s = \left[ \begin{array}{cccc}
     L_s^{(a_1)} &             &        &             \\
                 & L_s^{(a_2)} &        &             \\
                 &             & \ddots &             \\
                 &             &        & L_s^{(a_l)} 
                      \end{array} \right],
\label{generalsol}
\eeq
where $L_s^{(a)}$ is a generator of $a$-dimensional 
irreducible representation of $SU(2),$ and $a_1+a_2+\cdots+a_l=N.$ 
We remark that all the solutions of the form (\ref{generalsol}) do not 
contribute to the partition function, but only the $a_1=a_2=\cdots=a_l$ 
case does. In fact, considering the $l=2$ case, it is easy to see 
that the unbroken gauge group of the solution (\ref{generalsol}) 
contains a following $U(1)$-generator 
\beq
\left[ \begin{array}{cc}
                 e_1{\bf 1}_{a_1} & 0        \\
                         0        & e_2{\bf 1}_{a_2}
                \end{array} \right],
\eeq
where $e_1$ and $e_2$ are real parameters satisfying $e_1a_1+e_2a_2=0.$ 
Let us consider the $a_1\neq a_2$ case. 
The interactions are all in the form of commutators, 
so this $U(1)$ mode 
is free and massless, which implies that there exists 
a fermion zero-mode corresponding to the $U(1).$ Thus its contribution 
to the partition function vanishes. However, when $a_1=a_2,$ 
the $B_s$ is written as 
\beq
\frac im B_s={\bf 1}_2\otimes L_s^{(a)}, 
\eeq
where we put $a_1=a_2\equiv a.$ 
Then it is noted that 
the unbroken gauge group is enhanced to $SU(2)\otimes {\bf Z}_a.$ 
This mode is not free and thus can give a certain contribution to 
the partition function, which amounts to a 
$\cN=1$ $SU(2)\otimes {\bf Z}_a$ 
SYM theory. For the case of generic $l,$ a similar argument goes on 
and it can be seen that the vacuum which can contribute to the 
partition function is a $\cN=1$ $SU(l)\otimes {\bf Z}_a$ SYM theory 
with $al=N.$ In the consequence the partition function of 
the $\cN=4$ theory is represented as a sum of the partition function 
of the various $\cN=1$ theories: 
\beq
Z_{SU(N)}^{D=4,\; \cN=4}(T^4)=
\sum_{al=N}Z_{SU(l)\otimes {\bf Z}_a}^{D=4,\;\cN=1}(T^4).
\label{N=4partitionfunction1}
\eeq
Further, since the ${\bf Z}_a$ factor in the gauge group 
$SU(l)\otimes {\bf Z}_a$ implies a summation of the flat ${\bf Z}_a$ 
bundle, it yields a factor $a^{4-1}=a^3,$ 
where the power 4 comes from summing up the flat bundle for 
each $A_m$'s and the ($-1$) from dividing by the gauge group ${\bf Z}_a.$ 
Thus we have 
\beq
Z_{SU(N)}^{D=4,\; \cN=4}(T^4)=
\sum_{al=N}a^3Z_{SU(l)}^{D=4,\; \cN=1}(T^4).
\label{N=4partitionfunction2}
\eeq
Here we should note that the partition function of the $\cN=1$ theory 
in the right hand side 
is given by its Witten index. In the Hamiltonian formalism, 
the partition function is written as 
\beq
\Tr(-1)^Fe^{-\beta H}, 
\label{trform}
\eeq
where the $(-1)^F$-factor is included 
in order to impose the periodic boundary condition on the fermion fields. 
Recall that there is 
no Higgs field in the $\cN=1$ theory, 
which implies that there exists no continuous zero-mode 
in the theory on $T^4.$ 
Spectra appearing in the theory are all discrete.  
Then in eq. (\ref{trform}), the contribution of 
every supersymmetric pair with non-zero 
energy is precisely cancelled, and only the zero-energy states 
can contribute. 
So eq. (\ref{trform}) is independent of $\beta,$ and it coincides 
with the Witten index 
\beq
I\equiv\lim_{\beta\limit\infty}\Tr(-1)^Fe^{-\beta H}, 
\eeq
whose value of the $\cN=1$ $SU(l)$ SYM theory is known to be $l$ 
\cite{Witten5}. Plugging the above formulas, we obtain the answer 
\beq
Z_{SU(N)}^{D=4,\; \cN=4}(T^4)=\sum_{al=N}a^3l=N\sum_{a|N}a^2, 
\eeq
where the summation of $a$ in the right hand side is taken over 
the divisors of $N.$

 Also, for the gauge group $SU(N)/{\bf Z}_N$ we can obtain the 
partition function of the sector of zero 't Hooft magnetic flux 
applying the above argument to the ${\bf Z}_N$-factor 
\beq
Z_{SU(N)/{\bf Z}_N}^{D=4,\; \cN=4}(T^4)=
\frac{1}{N^3}Z_{SU(N)}^{D=4,\; \cN=4}(T^4)=
\sum_{a|N}\frac{1}{a^2}.
\eeq


\section{IIA Matrix String Partition Function}

   Here we calculate the partition function of IIA matrix string 
theory and compare the result of Kostov and Vanhove \cite{Kostov-Vanhove} 
which has been derived in the case of the strong coupling limit 
$g\limit\infty.$  
We have seen in section 2 that the SYM fields in periodic boundary 
condition can reproduce the second quantized superstrings with 
interactions. Thus, considering the two-dimensional SYM theory with 
periodic boundary condition as the IIA matrix string theory, we evaluate 
the partition function of this theory. 

  Now we have to take care of the $U(1)$ part of the gauge group $U(N)$ 
before doing the calculation. We consider the gauge group $U(N)$ 
in the factorized form 
\beq
U(N)=U(1)\times (SU(N)/{\bf Z}_N). 
\label{decomposeU(1)}
\eeq
The meaning of the $U(1)$ part in the IIA matrix string theory, 
whose field contents are $(A_{\mu},X,\theta),$ is as follows. 
The $U(1)$ part of $X$ and $\theta$ represents the center of mass 
coordinates of the strings in transverse directions. We fix it from 
the translational invariance. On the other hand, the $U(1)$ part of 
$A_{\mu}$ is related to the number of D-particles, so we have to take 
into account this. More precisely, as is discussed 
in \cite{DVV}, the $U(1)$ electric flux corresponds to the number of 
D-particles: 
\beq
q=\frac{1}{2\pi}\int_0^Rd\sigma E^{U(1)}\in {\bf Z}. 
\eeq
Corresponding to eq. (\ref{decomposeU(1)}), the gauge field $A_{\mu}$ 
is decomposed as 
\beq
A_{\mu}=A^{U(1)}_{\mu}T^{U(1)}+A_{\mu}^aT^a, 
\eeq
where $T^{U(1)}=\frac 1N{\bf 1}_N,$ and $T^a$ ($a=1,\cdots,N^2-1$) is a 
generator of $SU(N).$ Since the interaction in the SYM theory 
appears in the form 
of commutators, the $U(1)$ gauge part decouples and thus the 
partition function of the IIA matrix string theory becomes 
\beq
Z_{{\rm IIA-MS}}=\left(\int 
\frac{{\cal D}A^{U(1)}_{\mu}}{{\rm Vol}(U(1))}
e^{-S^{U(1)}}\right)Z_{SU(N)/{\bf Z}_N}^{D=2,\;\cN=8}(T^2), 
\label{ZIIAMS}
\eeq
where the Wick rotation was performed. The two-torus, 
where the SYM theory is defined on, is a rectangular one with the size 
$T\times R.$ The action of the $U(1)$ gauge part is given by 
$$
S^{U(1)}=\frac{1}{Ng^2}\int d^2\sigma\frac 14 
F^{U(1)}_{\mu\nu}F^{U(1)}_{\mu\nu}. 
$$
Let us evaluate the first factor in eq. (\ref{ZIIAMS}) by taking 
the $A^{U(1)}_0=0$ gauge. 
Then, the Gauss law constraint means that 
$A^{U(1)}_{\sigma}$ is independent of $\sigma.$ Also, considering the 
Wilson loop wrapping around the $\sigma$ direction, we see that 
$\theta(\tau)\equiv RA^{U(1)}_{\sigma}$ is an angular variable whose 
conjugate momentum $p$ is quantized to an integer. Employing the 
variable $\theta(\tau)$ and translating to the Hamiltonian form, 
we can compute the first factor as 
\begin{eqnarray}
\int \frac{{\cal D}A^{U(1)}_{\mu}}{{\rm Vol}(U(1))}\;e^{-S^{U(1)}} 
& = & \int_{\theta(T)=\theta(0)}{\cal D} \theta (\tau) 
       \;e^{-\frac 12 \frac{1}{Ng^2R}\int_0^Td\tau \dot{\theta}(\tau)^2} 
              \nn \\
& = & \Tr \left( e^{-T\frac 12 Ng^2Rp^2}\right)  \nn \\
& = & \sum_{p\in {\bf Z}}e^{-\frac{RT}{2}Ng^2p^2}. 
\label{firstfactor}
\end{eqnarray}

\subsection{$Z_{SU(N)/{\bf Z}_N}^{D=2,\;\cN=8}(T^2)$}

   Now our remaining task is evaluating the second factor of 
eq. (\ref{ZIIAMS}), which is performed by considering the dimensionally 
reduced version of the analysis in section 3\footnote{
Since in section 3 the supersymmetry algebra (\ref{***}) has no central 
charges, here we are to consider the $\cN=8$ supersymmetry algebra 
without central charges as the result of the dimensional reduction. 
In fact, the central charges are written in the form of total derivatives, 
so they do not appear under the periodic boundary condition. 
In the infinite two-dimensional space, central charges exist and 
represent the topological charges which characterize stable solitonic 
modes in the theory. On the contrary, in the two-torus 
with the finite size, such solitonic modes do not exist stably. 
In this case, it can be considered that with respect to the sector of 
the zero total charge, dynamical degrees of freedom of the modes are 
contained in the theory, i.e. in the configurations 
which the path integration is performed over, 
and that they arise as metastable states 
when the sizes of the two-torus becomes large enough. 
Of course, a similar consideration is possible in the three-torus case 
in section 5. }. 
Since the $Q$-exact 
structure as well as the $Q$-invariant one is preserved  
even after the dimensional reduction, 
the dimensionally reduced theory of a cohomological field theory 
becomes also a cohomological field theory. Further, the $U(1)$-symmetry 
in (\ref{U(1)charge}) remains nonanomalous in the dimensional reduction 
to two-dimensions in the 
case of the $SU(N)$ gauge group, because it contains no $U(1)$ factor. 
Thus, the arguments of the 
mass perturbation in section 3.2 can be applied also to the 
dimensionally reduced case. Then, the mass perturbation breaks 
the $\cN=8$ supersymmetry to $\cN=2,$ 
and the partition function takes the 
form 
\beq
Z_{SU(N)}^{D=2, \;\cN=8}(T^2)=
\sum_{al=N}Z_{SU(l)\otimes {\bf Z}_a}^{D=2, \;\cN=2}(T^2).
\label{D=2N=8partitionfunction1}
\eeq 
In two-dimensions, by the argument similar as in the four-dimensional 
case, the ${\bf Z}_a$ factor of the gauge group yields 
$a^{2-1}=a.$ Hence we have 
\beq
Z_{SU(N)}^{D=2,\; \cN=8}(T^2)=
\sum_{al=N}aZ_{SU(l)}^{D=2,\; \cN=2}(T^2).
\label{D=2N=8partitionfunction2}
\eeq 
Here the $\cN=2$ theory in two dimensions contains Higgs fields, 
whose zero momentum modes form continuous spectrum beginning with 
zero-energy because 
the field space of the Higgs fields is noncompact. This situation 
makes ambiguous the relation between the partition function 
and the Witten index, so we cannot go along the same line as 
in the four-dimensional case. 
However, owing to the fact that $\cN=2$ $SU(l)$ SYM theory 
in two dimensions is a cohomological field theory, 
the partition function 
$Z_{SU(l)}^{D=2,\; \cN=2}(T^2)$ can be evaluated. 
Let us see it from now. 

    After an appropriate field redefinition (see Appendix A), 
the classical action of the 
$\cN=2$ SYM theory is written as the BRST exact form 
\beq
S=Q\int d^2\sigma\;\tr\left(\frac{1}{8g^2}\eta[\phi,\bar{\phi}]
    -i\chi\Phi+2g^2\chi H+\frac{1}{2g^2}\psi_{\mu}D_{\mu}\bar{\phi}
     \right), 
\label{N=2TFT}
\eeq
where the BRST transformation is defined by 
\beq
\begin{array}{lll}
QA_{\mu}=\psi_{\mu}, & Q\psi_{\mu}=-iD_{\mu}\phi, & Q\phi=0,\\
Q\chi=H, & QH=[\phi, \chi], & \\
Q\bar{\phi}=\eta,& Q\eta=[\phi, \bar{\phi}]. & 
\end{array}
\eeq
Note that the $Q$ is nilpotent up to the gauge transformation with the 
parameter $\phi,$ which gives a cohomology to equivalent classes with 
respect to the gauge transformation. The field contents are as follows. 
$A_{\mu}$ is a two-dimensional gauge field, and $\psi_{\mu},$ $\eta,$ 
$\chi$ together stand for fermions. $\phi$ and $\bar{\phi}$ are 
complex Higgs fields, and $H$ is a bosonic auxiliary field. 
Ghost number is assigned as $-2$ to $\bar{\phi},$ $-1$ to 
$\eta$ and $\chi,$ $0$ to $A_{\mu}$ and $H,$ +1 to $\psi_{\mu},$ +2 
to $\phi.$ The ghost number conservation is nonanomalous for the same 
reason as in the case of the dimensional reduction of the $U(1)$-symmetry. 
The contribution of the path integration of the gauge field 
localizes in the configurations determined by 
\beq
\Phi\equiv -2F_{12}=0.
\label{Phiequation}
\eeq
The addition of a $Q$-exact term to the action does not 
change the theory, if it behaves well at infinity in the field space. 
Thus, we may discard the first term in eq. (\ref{N=2TFT}) 
\beq
Q\int d^2\sigma\;\tr\frac{1}{8g^2}\eta[\phi,\bar{\phi}]. 
\label{firstterm}
\eeq
Also, it can be seen that 
the partition function is independent of the coupling $g$ for the same 
reason.  
The localization (\ref{Phiequation}) 
can be shown by integrating out $H$ and $\chi$ fields in the 
$g\limit 0$ limit after the integrals of $\bar{\phi}$ and $\eta.$ 
Then the partition function becomes 
\beq
\int\frac{{\cal D}A_{\mu}}{{\rm Vol}(SU(l))}
{\cal D}\phi{\cal D}\psi_{\mu}
      \;\delta(D_{\mu}D_{\mu}\phi+\{\psi_{\mu},\psi_{\mu}\})
       \;\delta(D_{\mu}\psi_{\mu}) 
       \;\delta(D_1\psi_2-D_2\psi_1)\;\delta(F_{12}), 
\label{localization}
\eeq
which indicates the declared localization. 
It should be noted that the localization is determined by the BRST fixed 
point $Q\chi=0$ after using the equation of motion of $H.$ 
Eq. (\ref{localization}) is 
not in a suitable form for our purpose, so we will deform the theory 
judiciously as in the section 3 in ref. \cite{Witten6}. 

   We consider the action with the addition of the $Q$-exact term 
\beq
S(t)=S'+tQ\int d^2\sigma\;\tr\chi\bar{\phi}, 
\eeq 
where $S'$ stands for the action $S$ with the term (\ref{firstterm}) 
eliminated and $t$ is a parameter. 
If new BRST fixed points that flow in from infinity when $t$ turns on 
do not contribute, the deformed theory coincides with the original one. 
After integrating out $H,$ $\eta,$ $\chi$ and $\bar{\phi},$ we end up 
with the action in the large $t$ case 
\beq
S(t)= \frac{1}{2g^2t}\int d^2\sigma\;\tr
[F_{12}(D_{\mu}D_{\mu}\phi+\{\psi_{\mu},\psi_{\mu}\})
-iD_{\mu}\psi_{\mu}(D_1\psi_2-D_2\psi_1)]+O(t^{-2}), 
\label{S(t)1}
\eeq
which can be again written in the $Q$-exact form 
\beq
S(t)=Q\left[\frac{i}{2g^2t} \int d^2\sigma\;\tr 
F_{12}D_{\mu}\psi_{\mu}+O(t^{-2})\right]. 
\label{S(t)2}
\eeq
Considering the case $t=-iu$ with $u$ large real positive, 
the $\phi$-integration yields $\delta (D_{\mu}D_{\mu}F_{12}).$ 
Using the normalizability of $F_{12},$ it means that the localization 
realizes at the solutions of $D_{\mu}F_{12}=0,$ which contain 
extra components adding to the localization 
point of the original theory $F_{12}=0.$ Arising of the extra components 
is a signal of the flow of new BRST fixed points from infinity. 
In fact, in the above process, the contribution of the $\chi$ and 
$\bar{\phi}$ integrals localizes the points 
\beas
\chi & = & \frac{1}{2t}D_{\mu}\psi_{\mu}, \\
\bar{\phi} & = & -\frac 1t F_{12}+\frac{1}{2t^2}(D_{\mu}D_{\mu}\phi
         +\{\psi_{\mu},\psi_{\mu}\}), 
\eeas
which appear first when $t$ turns on. 
So they are the new fixed points flowing in from infinity. 
Thus in general, 
the deformed cohomological field theory does not equivalent 
to the original one. However, there is a possibility that the 
BRST invariant operators with the following feature exist --- 
in calculation of their correlators the extra components do not 
contribute. If there are such operators, the deformed theory coincides 
to the original one with respect to the restricted set of the operators. 
Indeed, we can find such operators. For the following BRST invariant 
operators 
\begin{eqnarray}
 & & \omega\equiv \int d^2\sigma \; \tr \;(-i\phi F_{12}+\psi_1\psi_2), \\
 & & \beta (\phi) = \tr\;(\;{\rm polynomial\; of \;}\phi), \nn 
\end{eqnarray}
we consider the unnormalized expectation value in the deformed theory 
\beq
\left\bra e^{\omega}\beta(\phi)\right\ket'\equiv
\int\frac{{\cal D}A_{\mu}}{{\rm Vol}(SU(l))}{\cal D}\phi{\cal D}\psi_{\mu}
\;\beta(\phi)\;e^{-S(-iu)+\omega}.
\label{expectationvalue}
\eeq
Here due to the $e^{\omega}$ factor, we can take the limit $u=\infty$ 
without changing the behavior of the fields at the infinity, 
and thus the limit 
does not change the value of (\ref{expectationvalue}). 
Further integrating out $\phi$ and $\psi_{\mu},$ we end up with 
\beq
\left\bra e^{\omega}\beta(\phi)\right\ket'=
\int\frac{{\cal D}A_{\mu}}{{\rm Vol}(SU(l))}\;
\beta\left(-i\frac{\delta}{\delta F_{12}}\right)\;\delta(F_{12}), 
\label{expectationvalue2}
\eeq
where the factor $\delta(F_{12})$ (with a finite degree of the 
derivative $\frac{\delta}{\delta F_{12}}$) 
indicates no contributions of the 
extra components. Therefore, $\left\bra e^{\omega}\beta(\phi)\right\ket'$ 
coincides with the unnormalized expectation value in the original theory 
which we denote by $\left\bra e^{\omega}\beta(\phi)\right\ket.$ 

   Now we can manage to compute the partition function 
$Z_{SU(l)}^{D=2,\; \cN=2}(T^2)=\bra 1\ket.$ Notice 
that $\omega$ has the ghost number $+2,$ and thus from the ghost number 
conservation we can show 
$\bra 1\ket=\bra e^{\omega}\ket.$ This is equal to the $\beta=1$ case 
of eq. (\ref{expectationvalue2}), so we find 
\beq
Z_{SU(l)}^{D=2,\; \cN=2}(T^2)=\int
\frac{{\cal D}A_{\mu}}{{\rm Vol}(SU(l))}\;
\delta(F_{12}), 
\label{partitionfunction3}
\eeq
which counts the number of the small gauge inequivalent configurations 
satisfying $F_{12}=0.$ The number is unity, 
since the two-dimensional $SU(l)$ gauge theory 
has no nontrivial winding number. 
In the consequence we have 
\beq
Z_{SU(l)}^{D=2,\; \cN=2}(T^2)=1. 
\eeq
It can be confirmed also 
by doing the concrete calculation, for example, employing the $A_1=0$ 
gauge fixing in eq. (\ref{partitionfunction3}). 
Plugging this into eq.(\ref{D=2N=8partitionfunction2}) we get the 
result 
\beq
Z_{SU(N)}^{D=2,\; \cN=8}(T^2)=\sum_{al=N}a. 
\eeq
Finally by taking account into the ${\bf Z}_N$ factor 
as in the four dimensional case, the second factor 
$Z_{SU(N)/{\bf Z}_N}^{D=2,\;\cN=8}(T^2)$ of eq. (\ref{ZIIAMS}) 
is obtained as 
\beq
Z_{SU(N)/{\bf Z}_N}^{D=2,\;\cN=8}(T^2)=
\frac 1N Z_{SU(N)}^{D=2,\; \cN=8}(T^2)
 =\sum_{a|N}\frac 1a.
\label{resultsecondfactor}
\eeq

\subsection{Result of IIA Matrix String Partition Function}

   Now we can write down the result of the partition function of 
the IIA matrix string theory. Substituting eqs. (\ref{firstfactor}) 
and (\ref{resultsecondfactor}) into eq. (\ref{ZIIAMS}), we have 
\beq
Z_{{\rm IIA-MS}}=\left(\sum_{a|N}\frac 1a\right)\sum_{p\in{\bf Z}}
e^{-\frac{RT}{2}Ng^2p^2},
\label{IIAresult1}
\eeq
which coincides the result obtained in the strongly coupled limit 
($g^2\limit \infty$) by Kostov and Vanhove \cite{Kostov-Vanhove}. 
They have conjectured that their result holds irrespectively of the 
strength of the coupling, and they called this property {\it exact 
quasi classics.} Since our calculation has been exactly performed 
without any approximation, it gives a proof of the {\it exact 
quasi classics.}

  Recalling the relations (\ref{RLrelation}) and (\ref{IIAggs}), 
we rewrite the result (\ref{IIAresult1}) 
in variables in string theory 
\beq
Z_{{\rm IIA-MS}}=\left(\sum_{a|N}\frac 1a\right)\sum_{p\in{\bf Z}}
e^{-T\frac{NL}{4\pi}\frac{p^2}{(g_sl_s)^2}}. 
\label{IIAresult2}
\eeq
There are two comments in order. 
First, the second factor in (\ref{IIAresult2}) 
represents a certain nonperturbative effect, which corresponds to 
creation and annihilation of D-particle and anti-D-particle pairs. 
Such phenomena as creation/annihilation of D- and anti-D- objects 
cannot be seen in the M(atrix) theory, 
because in the infinite momentum 
frame anti-D-particles 
in the M(atrix) theory are integrated out and do not appear. 
It can be seen first after compactified to the IIA matrix string theory. 
Second, there is no perturbative correction in the formula 
(\ref{IIAresult2}). It agrees to nonrenormalization theorems in 
perturbative superstring theory by Martinec \cite{Martinec}, 
which tells that the 0-, 1-, 2-, and 3-point functions of massless 
string vertex operators receive no perturbative corrections in 
flat ten-dimensional backgrounds.


\section{IIB Matrix String Partition Function}

   Here, we compute the partition function of IIB matrix string 
theory in the ten-dimensional IIB limit, and compare the exact result 
of the IKKT model by 
Moore-Nekrasov-Shatashvili \cite{Moore-Nekrasov-Shatashvili}. 
We consider the three-dimensional SYM theory with periodic boundary 
condition as the IIB matrix string theory for the same reason 
as in the 
IIA case. 
With respect to the gauge group, the $U(1)$ part of the gauge group 
$U(N)=U(1)\times (SU(N)/{\bf Z}_N)$ 
for the field contents $(A_{\mu}, X, \theta)$ corresponds to 
the center of mass coordinates of the strings in transverse 
directions in the ten-dimensional IIB limit ``$R_1R_2\limit\infty$ 
with $g_s$ fixed,'' where the $U(1)$ part of the gauge field together 
with that of $X$ become the transverse coordinates. Thus we fix 
the $U(1)$ part of the gauge field as well as the Higgs fields, 
and consider the partition function of the 
three-dimensional $\cN=8,$ $SU(N)/{\bf Z}_N$ SYM theory as that of 
the IIB matrix string theory 
\beq
Z_{{\rm IIB-MS}}=Z_{SU(N)/{\bf Z}_N}^{D=3, \; \cN=8}(T^3).
\label{ZIIBMS}
\eeq
The three-torus, where the SYM theory is defined on, is taken 
to be rectangular of the size $T\times R_1 \times R_2,$ with the 
Euclidean signature. 
(We performed the Wick rotation as in the IIA case.) 
In this case, we can also use the dimensionally reduced version of 
the arguments of the mass perturbation in the four dimensions, 
due to the following two reasons. One is that the dimensional reduction 
of a cohomological field theory is also a cohomological field theory. 
The other 
is that in odd dimensions 
the dimensionally reduced version of the $U(1)$-symmetry 
in eq. (\ref{U(1)charge}) is anomaly free. 

  Going on along the same line as in the IIA case, we have 
\begin{eqnarray}
Z_{SU(N)}^{D=3,\; \cN=8}(T^3) & = & 
\sum_{al=N}Z_{SU(l)\otimes{\bf Z}_a}^{D=3, \; \cN=2}(T^3) \nn \\
 & = & \sum_{al=N}a^2Z_{SU(l)}^{D=3, \; \cN=2}(T^3).
\label{ZD=3N=8SYM}
\end{eqnarray}
Here, there exists a Higgs field in the three-dimensional $\cN=2$ 
theory. For the same reason as before, we cannot relate 
the partition function 
directly to the Witten index. However, if considering 
the ten-dimensional IIB limit (\ref{10DIIBlimit}) 
which in fact we are interested in, we can proceed further. 
Note that in the large volume limit physics becomes independent of 
the boundary condition. So we can evaluate 
$Z_{SU(l)}^{D=3, \; \cN=2}(T^3)$ 
by adopting a twisted boundary condition instead of the periodic 
boundary condition.

\subsection{$\cN=2$ partition function with a twisted boundary 
condition}

   Here we consider the three-dimensional $\cN=2$ $SU(l)$ SYM theory 
with the twisted boundary condition 
\begin{eqnarray}
A_i(t,\sigma_1,\sigma_2) & = & A_i(t+T, \sigma_1,\sigma_2) \nn \\
    & = & PA_i(t,\sigma_1+R_1, \sigma_2)P^{-1} \nn \\
    & = & QA_i(t,\sigma_1, \sigma_2+R_2)Q^{-1}, 
\label{twistedBC}
\end{eqnarray}
where $i=1,2,$ we took the $A_0=0$ gauge fixing, and $P$ and $Q$ are 
$SU(l)$ matrices satisfying $PQ=QPe^{2\pi i/l}.$ For example, $P$ 
and $Q$ can be represented as 
\beq
P=e^{-\pi i\frac{l+1}{l}}\left[\begin{array}{ccccc}
   0 & 1 &       &        &   \\
     & 0 & 1     &        &   \\
     &   &\ddots & \ddots &   \\
     &   &       & \ddots & 1 \\
   1 &   &       &        & 0  
   \end{array}\right],
\hspace{1cm}
Q=e^{-\pi i\frac{l-1}{l}}\left[\begin{array}{ccccc}
   1 &             &              &       &     \\
     & e^{2\pi i/l}&              &       &     \\
     &             & e^{4\pi i/l} &       &     \\
     &             &              & \ddots&     \\
     &             &              &       & e^{2\pi i(l-1)/l}
   \end{array}\right].
\eeq
Supersymmetry requires that the other fields (Higgs $\phi$ and complex 
fermion $\lambda$) satisfy the same boundary condition 
as eq. (\ref{twistedBC}). Under this 
boundary condition, zero momentum modes become trivial 
\beq
A_i^{(0)}=\phi^{(0)}=\lambda^{(0)}=0, 
\label{****}
\eeq
because the constant traceless hermitian matrices 
commuting with 
$P$ and $Q$ simultaneously do not exist 
except the trivial case (\ref{****}). 
 
  In this situation, spectra appearing in the theory are discrete, 
and thus the partition function coincides the Witten index. 
Let us consider the Witten index. The argument below is 
a three-dimensional analogue of Witten's consideration 
in the four-dimensional case \cite{Witten6}.  Also, though it is 
briefly reported in \cite{Affleck-Harvey-Witten}, we will discuss it 
in order to make this paper more self-contained. 
Our problem is now 
reduced to counting the number of vacua, i.e. the number 
of (small) gauge inequivalent classes of the classical zero energy 
states.  
The gauge transformation preserving both of the boundary condition 
(\ref{twistedBC}) and the gauge condition $A_0=0$ is generated by 
the time-independent 
$SU(l)$ matrix of the  following boundary condition 
\begin{eqnarray}
U(\sigma_1,\sigma_2) & = & 
e^{2\pi ik_1/l}PU(\sigma_1+R_1,\sigma_2)P^{-1} \nn \\
 & = & e^{2\pi ik_2/l}QU(\sigma_1,\sigma_2+R_2)Q^{-1} 
\label{UBC}
\end{eqnarray}
where $k_i=0,1,\cdots,l-1.$ 
The classical zero energy state is given by the configuration 
$$
A_i=-i(\partial_iU)U^{-1}
$$
with the other fields nil. Here, if this $U$ can be continuously deformed 
to the identity, there are no nontrivial sectors, and thus the vacuum 
is unique modulo small gauge transformations. We will see that 
it is in fact so. The $U$ can be written by the $SU(l)$ matrix 
$\tilde{U}$ satisfying the simpler boundary condition 
\beq
U=(Q^{-1})^{k_1}P^{k_2}\tilde{U}, 
\label{UUtilde}
\eeq
where 
\begin{eqnarray}
\tilde{U}(\sigma_1,\sigma_2) & = & 
P\tilde{U}(\sigma_1+R_1,\sigma_2)P^{-1} \nn \\
 & = & Q\tilde{U}(\sigma_1,\sigma_2+R_2)Q^{-1}. 
\label{UtildeBC}
\end{eqnarray}
We consider a topological classification of gauge transformations 
with the boundary condition (\ref{UtildeBC}). 
Because of $\pi_0(SU(l))=0,$ 
by a suitable continuous translation in a group manifold of $SU(l),$ 
we can always 
start with $\tilde{U}(0,0)=1.$ Then, using eq. (\ref{UtildeBC}) 
we see that 
$$
\tilde{U}(0,0)=\tilde{U}(R_1,0)=\tilde{U}(0,R_2)=\tilde{U}(R_1,R_2)=1, 
$$
i.e. $\tilde{U}$'s on the vertices of a square with the size 
$R_1\times R_2$ in $(\sigma_1, \sigma_2)$-space are all identity. From 
the fact $\pi_1(SU(l))=0,$ $\tilde{U}$ on the edges of the square can 
be taken identity by a continuous deformation. Further, using 
$\pi_2(SU(l))=0,$ we continuously deform to $\tilde{U}=1$ everywhere 
in the square. 
Using $\pi_0(SU(l))=0$ again, by a continuous translation 
the $U$ in eq. (\ref{UUtilde}) can be taken identity 
on the square. 
Thus, it is confirmed that there is no nontrivial 
topological vacuum sector. We conclude that the Witten index is unity, 
which leads to 
\beq
Z_{SU(l)}^{D=3, \; \cN=2}(T^3)|_{{\rm twisted \;B.C.}}=1. 
\label{ZtwistedBC}
\eeq

\subsection{Result of IIB Matrix String Partition Function}

   As discussed before, in the ten-dimensional IIB limit 
we can replace the value of the partition function 
$Z_{SU(l)}^{D=3, \; \cN=2}(T^3)$ with that of eq. (\ref{ZtwistedBC}) 
\beq
Z_{SU(l)}^{D=3, \; \cN=2}(T^3)=1. 
\label{ZperiodicBC}
\eeq
Substituting this into eq. (\ref{ZD=3N=8SYM}), we have 
\beq
Z_{SU(N)}^{D=3,\; \cN=8}(T^3) = \sum_{al=N}a^2,
\eeq
and thus the partition function of the IIB matrix string theory is 
obtained as 
\begin{eqnarray}
Z_{{\rm IIB-MS}} & = & Z_{SU(N)/{\bf Z}_N}^{D=3, \; \cN=8}(T^3) \nn \\
 & = & \frac{1}{N^2}Z_{SU(N)}^{D=3,\; \cN=8}(T^3) \nn \\
 & = & \sum_{a|N}\frac{1}{a^2},
\label{ZIIBMS2}
\end{eqnarray}
which is valid in the ten-dimensional IIB limit (\ref{10DIIBlimit}). 
We should remark that the result (\ref{ZIIBMS2}) 
coincides the result of the IKKT model \cite{IKKT} by Moore, Nekrasov 
and Shatashvili \cite{Moore-Nekrasov-Shatashvili}. 
It is quite nontrivial and might be a signal of the equivalence 
between the two IIB matrix models arising from the different roots 
--- one is a compactification of the BFSS M(atrix) theory, 
and the other is a 
matrix regularization of the worldsheet action of type IIB superstring 
in Schild gauge. 
Also, the result (\ref{ZIIBMS2}) has neither perturbative nor 
nonperturbative correction. With respect to the former it agrees again 
with the nonrenormalization theorems in perturbative superstring theory 
\cite{Martinec}.


\section{Conclusions}

   We have considered the IIA and IIB matrix string theories derived from 
the M(atrix) theory via toroidal compactification. We have summarized 
that string interactions emerge as the Yang-Mills instantons and 
have shown 
that in the IIB matrix string theory the chiral spinors are correctly 
reproduced. 

    As a preparation for computation of 
the matrix string partition functions, we have calculated the partition 
function of four-dimensional SYM theory on $T^4,$ by mapping 
the theory to a cohomological field theory. 
Here, considering the mass perturbation (\ref{masspert}) has made 
the calculation easier. We have shown the invariance of the partition 
function under the mass perturbation in a different fashion from 
the argument by Vafa and Witten \cite{Vafa-Witten}.  

   We have exactly computed the partition function of the IIA matrix 
string theory by mapping the theory into a cohomological field theory. 
Our result coincides with the result obtained in the infra-red limit 
by Kostov and Vanhove \cite{Kostov-Vanhove}, and thus gives a proof of 
the {\it exact quasi classics} of the theory conjectured by them. 
The formula for the partition function receives no perturbative 
correction, which is in conformity with nonrenormalization theorems 
in perturbative superstring theory by Martinec \cite{Martinec}. 
Also, there exist some nonperturbative corrections 
which come from the $U(1)$ electric flux in the SYM theory 
and which are interpreted 
as creation and annihilation of D-particle and anti-D-particle pairs. 
Such phenomena have been reported in high energy scattering of strings 
in ref. \cite{GHV}. It may be interesting to deepen the meaning of our 
result from the line of the high energy scattering. 

   Further, we have evaluated the partition function of the IIB matrix 
string theory in the ten-dimensional IIB limit by a similar method as 
in the IIA case. 
Our result receives neither perturbative and nonperturbative 
corrections, which with respect to the former agrees with the 
nonrenormalization theorems again. 
Also, our result coincides with the exact result of 
the partition function of the IKKT model by 
Moore, Nekrasov and Shatashvili \cite{Moore-Nekrasov-Shatashvili}. 
Although both of the IIB matrix string theory 
and the IKKT model are considered 
to give type IIB string theory, 
they have arised from the distinctive 
origins, and the relation between them have not been 
clarified yet. Thus our result is quite nontrivial, and may suggest 
the equivalence of those two models. 
In the IIB matrix string theory, the S-duality is well understood from 
a geometry of the two-torus, but the ten-dimensional Lorentz symmetry 
is not manifest. On the other hand, in the IKKT model side, 
while there is a 
manifest Lorentz symmetry, we have not been able to see the S-dual 
structure. 
In this situation, it seems to be an important step to explore the 
equivalence and to establish the precise correspondence between them 
toward constructing the nonperturbative definition which manifestly 
realizes both of 
the Lorentz symmetry and the S-dual structure.

\vspace{3cm}


\begin{large}
{\bf Acknowledgements}
\end{large}
\vspace{7mm}

   The preliminary version of this work was presented 
in KEK theory workshop '99. The author would like to thank the 
organizers and participants of the workshop, and especially 
N. Ishibashi, S. Iso, H. Kawai, T. Kuroki, Y. Okada, K. Okuyama 
and A. Tsuchiya for valuable conversations and encouragements. 
The research of the author is supported by the Japan Society for 
the Promotion of Science under the Postdoctral Research Program. 

\vspace{3cm}


\begin{large}
{\bf Note Added}
\end{large}
\vspace{7mm}

   While writing up the manuscript, I received the papers \cite{GS,P} 
which discuss issues close to this work. 
In \cite{GS} 
the thermodynamic partition function of the IIA matrix 
string theory is calculated in the $g_s\limit 0$ limit, 
and in \cite{P} some thermodynamic properties of 
the IIA and IIB matrix string theories are discussed.

\newpage

\begin{large}
{\bf Appendix}
\end{large} 

\appendix


\section{Two-Dimensional $\cN=2$ $SU(l)$ Super Yang-Mills Theory as a 
Cohomological Field Theory}

    Here we see that the cohomological field theory (\ref{N=2TFT}) is 
identical to Euclidean $\cN=2$ $SU(l)$ SYM theory in two dimensions. 
We perform the Wick rotation 
for four-dimensional $\cN=1$ $SU(l)$ SYM theory, 
and then see that its dimensional reduction to two dimensions coincides 
with eq. (\ref{N=2TFT}). 

    First, we start with $\cN=1$ $SU(l)$ SYM theory in Minkowskian 
four-dimensional space-time 
\beq
S_M=\frac{1}{g^2}\int d^4\sigma \;\tr \left(
-\frac 14 F_{\mu\nu}F^{\mu\nu}
+\frac i2 \bar{\Psi}\Gamma^{\mu}D_{\mu}\Psi\right), 
\eeq
where $\bar{\Psi}=\Psi^{T}\Gamma^0,$ our convention of the metric is 
$\eta^{\mu\nu}={\rm diag}(-1,+1,+1,+1),$ and the $\Gamma$-matrices satisfy 
$$
\{\Gamma^{\mu}, \Gamma^{\nu}\}=-2\eta^{\mu\nu}.
$$
After defining the $\gamma$-matrices as 
$\gamma_i=\gamma^i=-i\Gamma^0\Gamma^i$ ($i=1,2,3$) which satisfy 
\beq
\{\gamma^i, \gamma^j\}=-2\delta^{ij}, 
\label{A-2}
\eeq 
we perform the Wick rotation as $\sigma^0=-i\sigma^4.$ 
Then, we have the following Euclidean 
action 
\beq
S_E=\frac{1}{g^2}\int d^4\sigma\;\tr \left[
\frac 14 (F_{\mu\nu})^2+\frac 12 \Psi^T(\gamma_iD_i+D_4)\Psi\right]. 
\label{A-1}
\eeq

  On the other hand, in eq. (\ref{N=2TFT}), integrating out $H$ and 
reviving the gauge fields as 
$$
\phi=A_3+iA_4, \hspace{1cm} \bar{\phi}= A_3-iA_4, 
$$
it can be easily seen that the bosonic part of the action is 
the dimensional reduction of 
$$
\int d^4\sigma\;\tr\frac{1}{4g^2}(F_{\mu\nu})^2. 
$$
Also, after defining the spinor in four dimensions as 
\beq
\Psi\equiv\left(\begin{array}{c} \psi_1 \\ 
                                \psi_2 \\  
                              \frac 12 \eta \\
                             2g^2\chi 
                \end{array} \right), 
\eeq
the fermion part takes the form of the dimensional reduction of 
$$
\int d^4\sigma\;\tr\frac{1}{2g^2}\Psi^T(\gamma_iD_i+D_4)\Psi, 
$$
where $\gamma_i$'s are 
\beq
\gamma_1=-i\left[\begin{array}{cc} 0 & {\bf 1}_2 \\
                                 {\bf 1}_2 & 0 
                 \end{array} \right],
\hspace{1cm}
\gamma_2=\left[\begin{array}{cc} 0 & -\sigma_y \\
                                 \sigma_y & 0 
                 \end{array} \right],
\hspace{1cm}
\gamma_3=i\left[\begin{array}{cc} {\bf 1}_2 & 0 \\
                                      0 & -{\bf 1}_2 
                 \end{array} \right],
\eeq
and satisfy eq. (\ref{A-2}). 
Thus it shows that the cohomological field theory (\ref{N=2TFT}) coincides 
with the dimensional reduction of (\ref{A-1}) to two dimensions.

\newpage
 

\end{document}